# Quantum metrology enhanced by repetitive quantum error correction

Thomas Unden[1], Priya Balasubramanian[1], Daniel Louzon[1,3], Yuval Vinkler[3], Martin B. Plenio[2,4], Matthew Markham[5], Daniel Twitchen[5], Igor Lovchinsky[6], Alexander O. Sushkov[6], Mikhail D. Lukin[6], Alex Retzker[3], Boris Naydenov[1,2], Liam P. McGuinness[1], Fedor Jelezko[1,2]*

[1] Institute of Quantum Optics, Ulm University, 89081 Ulm, Germany

[2] Center for Integrated Quantum Science and Technology (IQST), Ulm University, 89081 Ulm, Germany

[3] Racah Institute of Physics, Hebrew University of Jerusalem, 91904 Jerusalem, Israel

[4] Institute for Theoretical Physics, Ulm University, 89081 Ulm, Germany

[5] Element Six, Harwell Campus, Fermi Avenue, Didcot, OX11 0QR, United Kingdom

[6] Department of Physics, Harvard University, Cambridge, Massachusetts 02138, USA

* E-Mail: fedor.jelezko@uni-ulm.de

The accumulation of quantum phase in response to a signal is the central mechanism of quantum sensing, as such, loss of phase information presents a fundamental limitation. For this reason approaches to extend quantum coherence in the presence of noise are actively being explored. Here we experimentally protect a room-temperature hybrid spin register against environmental decoherence by performing repeated quantum error correction whilst maintaining sensitivity to signal fields. We use a long-lived nuclear spin to correct multiple phase errors on a sensitive electron spin in diamond and realize magnetic field sensing beyond the timescales set by natural decoherence. The universal extension of sensing time, robust to noise at any frequency, demonstrates the definitive advantage entangled multi-qubit systems provide for quantum sensing and offers an important complement to quantum control techniques. In particular, our work opens the door for detecting minute signals in the presence of high frequency noise, where standard protocols reach their limits.

Precise magnetic and electric field sensors constructed from qubits use the quantum phase of coherent superposition states to acquire a measurable signal. The coherence time $T_2$, of the qubit state therefore imposes an intrinsic limit on sensing techniques, with sensitivity scaling as $1/\sqrt{T_2}$ [1]. Following this principle, techniques to extend quantum coherence in different systems have been developed to enable the detection of weaker signals with greater precision. However, conventional methods such as environmental purification and cooling are now reaching their technical limits. Quantum control in particular has proved exceptional at extending quantum coherence by decoupling the sensor from the environment in selected frequency bandwidths [2-4]. Yet, while central to enhancing quantum metrology in noisy environments, dynamical decoupling fails in the presence of high-frequency noise with short correlation time, and is ultimately restricted by $T_1$ relaxation.

Recently, quantum error correction has been proposed to improve quantum metrology in the presence of noise [5-10]. Rather than driving the system faster than the typical noise frequency, as occurs in dynamical decoupling, interaction of the sensing qubit with the signal and environment remains unperturbed. Additional qubits then allow unwanted entropy introduced from the environment to be selectively extracted through quantum logic operations. By virtue of placing no restriction on the frequency spectrum of environmental noise, error correction operates when the sensor is subjected to fast noise and thus constitutes a promising alternative to state-of-the-art quantum sensing techniques. In the context of quantum information processing, quantum error correction and coherent feedback with a nuclear spin quantum register was recently demonstrated with nitrogen-vacancy (NV) centers [11-13] and several other quantum systems [14-16]. In what follows we apply quantum error correction to enhance the sensitivity of quantum metrology under ambient conditions. We show full control over the register including multiple resets of sensing qubit without perturbing the robust qubit, allowing up to two rounds of repetitive quantum error correction to be performed. We subsequently apply error correction to improve the sensor's phase coherence time in the presence of natural errors that do not affect the robust qubit, whilst also demonstrating high frequency magnetic field sensing over extended time periods.

Our experiments are based on a hybrid quantum register formed by the electronic spin of an NV center and a nearby $^{13}$C nuclear spin in diamond that exhibit remarkably long coherence times even under ambient conditions [17] (Figure 1.A). Both spins can be coherently prepared with high fidelity and independently manipulated; using microwave (MW) radiation in the case of the electron $m_s = 0,-1$ states ($|0\rangle$, $|-1\rangle$), and radiofrequency (RF) radiation in the case of the nuclear $m_s = \pm\frac{1}{2}$ states ($|\uparrow\rangle$, $|\downarrow\rangle$). Magnetic dipolar interaction between the spins further allows for quantum gates to be performed (Figure 1.B). Use of a hybrid register where a robust qubit with long $T_2$ time and a sensing qubit which interacts strongly with the signal (but with significantly shorter $T_2$ time) are simultaneously available, allows a quantum error correction protocol specific for quantum metrology to be designed. In our protocol, which we term quantum sensing with error correction (QSEC), several noisy qubits are replaced by a single noiseless qubit, thereby requiring fewer resources than conventional majority voting with three qubit code. The robust qubit, insensitive to errors, is used to actively remove errors accumulated on the sensing qubit whilst leaving interaction with the signal unperturbed.

Our QSEC approach proceeds as follows. We define a code space spanned by $|+\rangle|\downarrow\rangle$ and $|-\rangle|\uparrow\rangle$ in which dynamics mediated by the signal takes place, where $|+\rangle = |0\rangle + i|-1\rangle$ and $|-\rangle = |0\rangle - i|-1\rangle$. The experiment starts with the generation of an entangled state $|+\rangle|\downarrow\rangle + |-\rangle|\uparrow\rangle$. Information about the signal field is encoded in the accumulation of a quantum phase $\Phi$, resulting in the state $|+\rangle|\downarrow\rangle + \exp(i\Phi)|-\rangle|\uparrow\rangle$ (see Figure 2.A). An NV phase-flip error ( $|+\rangle \leftrightarrow |-\rangle$) occurring on a state in the protected code space maps the system to an orthogonal error subspace generated by the states $|+\rangle|\uparrow\rangle$ and $|-\rangle|\downarrow\rangle$. In the error subspace, the state accumulates an additional phase $|-\rangle|\downarrow\rangle + \exp(i[\Phi + \Phi_E])|+\rangle|\uparrow\rangle$, which can be interpreted as a flip of the rotation axis of the Bloch vector. Without correction of such errors, the signal encoded in the phase of the quantum register undergoes a random walk, instead of increasing linearly with time. For simplicity, we assume there is no time evolution between the error occurrence and the following correction step in the scheme presented in Figure 2.A, i.e. . By using orthogonality

between the code space and the error subspace, errors can be detected and corrected to bring the system back to the code space.

Experimentally we achieve phase error correction by a three step process. First, a conditional $\pi$ rotation of the electron spin around the z-axis, dependent on the nuclear state is performed (CNOT gate), which maps the accumulated phase onto the nuclear spin (Decoding). After a $\pi/2$ -pulse is applied to the electron spin, an error occurrence is manifested in the state of the NV center, namely $|0\rangle$ when there was no error and $|-1\rangle$ when there was an error. Second, the sensor qubit is optically reset conditional on an error occurring, yielding $|0\rangle|\downarrow\rangle + \exp(i\Phi)|0\rangle|\uparrow\rangle$ (Reset). Finally, a second microwave pulse followed by a CNOT gate, returns the state to the code space $|+\rangle|\downarrow\rangle + \exp(i\Phi)|-\rangle|\uparrow\rangle$ (Encoding). In the case an error did not occurred, the final state, $|+\rangle|\downarrow\rangle + \exp(i\Phi)|-\rangle|\uparrow\rangle$ is identical to that which would be obtained without error correction, i.e. is unaltered by error correction.

We use this aspect of error correction, the reset of the sensor conditional on the occurrence of an error, to counter the non-unitary nature of dephasing. In particular, when the elapsed time between error and subsequent correction is much shorter than the precession period of the corresponding Bloch vector, the phase mismatch $\Delta\Phi = |\Phi_C - \Phi_E|$ is removed and sensing can continue with negligible phase information loss. Subsequent readout of the nuclear spin, performed by a non-demolition measurement, then ideally gives the same result regardless of errors affecting the sensing qubit. In order to be effective, each of the multiple steps of our QSEC procedure must be repetitively applied so that correction can be performed before errors destroy phase information. First, the signal is accumulated. Second, the accumulated signal is transferred to the nuclear spin while the entropy of the sensor is retained in the electron spin. Next, the entropy is removed by resetting the electron spin. In the final step, the electron and nuclear spin are correlated again. Such a repeated quantum error correction requires precise control of interactions in the quantum register and has only been realized for trapped ions [15].

The key element for the experiments presented in this work is the identification of a robust nuclear spin located along the axis of the NV center. In this unique physical geometry the dipolar Hamiltonian between the sensing and auxiliary qubit is secular, which allows two vital processes to be realized: *i)* precise initialization and readout of the quantum register and *ii)* optical reset of the NV spin while avoiding nuclear spin dephasing. The axial requirement may be relaxed in future experiments by selecting a weakly coupled nuclear spin and employing a high external magnetic field or RF-driving of the nuclear spin during optical reset.

We first investigated the ability to initialize, coherently control and readout the register. We obtained a single-shot readout fidelity of the nuclear spin of 98%, and an initialization fidelity of the register into the $|0\rangle|\uparrow\rangle$ state beyond the detection threshold. Quantum state tomography measurements of the entangled state $|+\rangle|\uparrow\rangle+|-\rangle|\downarrow\rangle$ gives a fidelity of 80% for the final state, which is limited primarily by gate errors (Figure 1.C). Coherence measurements of the nuclear spin under continuous optical reset of the electron spin verified the absence of nuclear dephasing for periods much longer than the reset time. For illumination periods exceeding 20 optical resets of the electronic spin, the nuclear coherence remained at about 95% of its initial value, with no decay observable (Figure 1.D).

To implement QSEC, we applied a weak magnetic field resonant with the NV spin, and recorded the acquired phase of the hybrid sensor. This scenario provides one example where QSEC has advantages over dynamical decoupling, since when the sensing field is resonant with the qubit energy scale, decoupling is impractical due to the requirement for pulse control on timescales shorter than the signal period. Of note, relaxometry techniques have recently been developed to exploit sensing of high frequency fields, with a precision limited by the sensor dephasing time [19, 20]. In Figure 2.A,B, we present our measurement sequence which incorporates two error correction repetitions into the sensing period, and is compared to a sequence without error correction. The acquired phase of the sensor leads to rotations on the Bloch sphere which are evidenced by coherent Rabi oscillations driven by the signal field (Figure 3.A). Without error correction, the oscillations decay on a timescale of 30 μs, limited by the dephasing time of the electron spin. After application of two rounds of error correction during the sensing period this limit is overcome and the sensing duration is significantly extended. In addition, a slower coherent oscillation is observable, which is due to the hyperfine interaction within the register (see simulation in Figure 3.C and [18]).

It is important to note that, due to our decoding sequence (see Figure 2.A), the accumulated signal is only converted to nuclear spin population when the electronic spin is in the $|0\rangle$ state: by application of a $\pi/2$-pulse. Consequently, our readout is designed to work only when there was no earlier error. The last error correction step therefore improves the contrast of the measurement while the intermediate correction prolongs the coherent interaction between sensor and signal.

Here, we emphasize that the corrected errors are environmental without any artificial applied noise source, and are treated naturally by our QSEC scheme. Both timing and amplitude of the errors are random, and are related to laboratory temperature and magnetic field fluctuations which act to limit the phase coherence time of the sensor. In our experiments, the signal field was switched off during QEC to allow for higher process fidelity, by using robust CNOT gates this requirement can be removed.

In addition, we demonstrate that correction of bit-flip errors with QSEC allows $T_1$ limits to quantum metrology to be in principle overcome. Here a non-resonant signal at 100 kHz was detected using a Carr-Purcell-Meiboom-Gill (CPMG) sequence and high frequency noise of varying strength was applied to the register to reduce the electron $T_1$ time from several milliseconds to as short as 60 ns [18]. As shown in supplemental Figure 7, without QSEC, the acquired signal is barely resolvable due to strong errors on the register, however when QSEC is implemented, a coherent signal with high contrast is observed despite the sensor spin experiencing multiple flips which would otherwise destroy coherence. As a result, the magnetic field sensitivity after application of QSEC could be increased by more than an order of magnitude. Furthermore, by concatenating both QSEC and CPMG sequences in this measurement, we integrate quantum error correction within a dynamical decoupling protocol to further exploit their complementarity. Indeed, in environments where low frequency noise is dominant, quantum control has proven effective at extending the sensor coherence to the $T_1$ limit, at which point QSEC may be employed to surpass relaxation.

To characterize the metrology improvements that were experimentally achieved in the presence of natural noise, we analyzed the data in Figure 3.A using a fitting algorithm based on Bayesian statistics [21]. By fitting the probability distribution function for the sensor decoherence rate γ, we obtain a 50% improvement in coherence time $T_2 = 1/\gamma$ when repetitive error correction is used, and accordingly a narrower linewidth which allows for improved signal discrimination (Figure 3.B). Simulations based on the system's master equation confirm this improvement of the coherence time (Figure 3.C). We also analyzed improvement in sensitivity $S$, when QSEC is used, by taking into account the contrast $C$ of the measured signal according to $S \propto \dfrac{1}{C\sqrt{T_2}}$ [1] (Figure 3.D). We find that the improvements obtained by an increase in sensing time are offset by a 30% lower contrast when QSEC is used due to imperfect gates, which also limit the number of error correction steps that can be performed. Consequently, we do not observe an absolute improvement in sensitivity. We do however, observe that at sensing times exceeding the intrinsic $T_2^*$ time of the sensing qubit, QSEC outperforms an uncorrected qubit for magnetic sensitivity. This is an important issue, since longer coherent interaction with the signal is required for optimized quantum sensing [22, 23], which benefits from Heisenberg scaling $S \propto \dfrac{1}{T}$.

By analyzing the fidelity of a single correction step we observe no limit due to nuclear spin dephasing caused by the optical NV reset (see Figure 1.D). Therefore a significantly increased performance is expected if robust gates insensitive to slow drifts of the NV center Larmor frequency are implemented.

Our results show that by using repetitive QEC, environmentally induced errors can be corrected to extend the sensing duration of quantum metrology, whilst also preserving sensitivity to signal fields. The combination of a robust and a sensitive qubit in QSEC shares a resemblance with the successful application of quantum logic for spectroscopy [24] in which a highly robust but hard to measure qubit is combined with a less robust but easy to detect qubit that reports on the quantum state of the robust qubit. In QSEC these roles are interchanged and, importantly, the role of the robust qubit in our scheme is the storage of phase information during error correction and extraction of entropy. Our demonstration of quantum error correction enhanced sensing protocol will be crucial for the rapidly growing field of diamond magnetometry, especially for NV assisted nanoscale magnetic resonance imaging and magnetic resonance spectroscopy. In these experiments high frequency noise originating from the surface is the main hurdle that has to be overcome [25]. Moreover, this technique can be extended to other quantum technologies including atomic clocks, NMR and fault-tolerant quantum computation. Robust control of several nuclear spins was demonstrated recently [12, 26] and promising (easily accessible) axial nuclear spins were identified [27]. The ability to manipulate these spins creates a promising platform for realizing more advanced QSEC codes that could tackle more complex noise models [9, 28]. The techniques demonstrated here will open a new field in quantum sensing, based on quantum error correction instead of, or in combination with dynamical decoupling and will have applications in various fields of science.

This work is supported by the EU (ERC, DIADEMS, SIQS, EQUAM), DFG (SFB/TR21 and FOR1993), DARPA, an Alexander von Humboldt Professorship and Volkswagenstiftung. We thank Michael Ferner and Manfred Bürzele for technical assistance, Paz London and Junichi Isoya for stimulating discussions.

**Figure 1.**

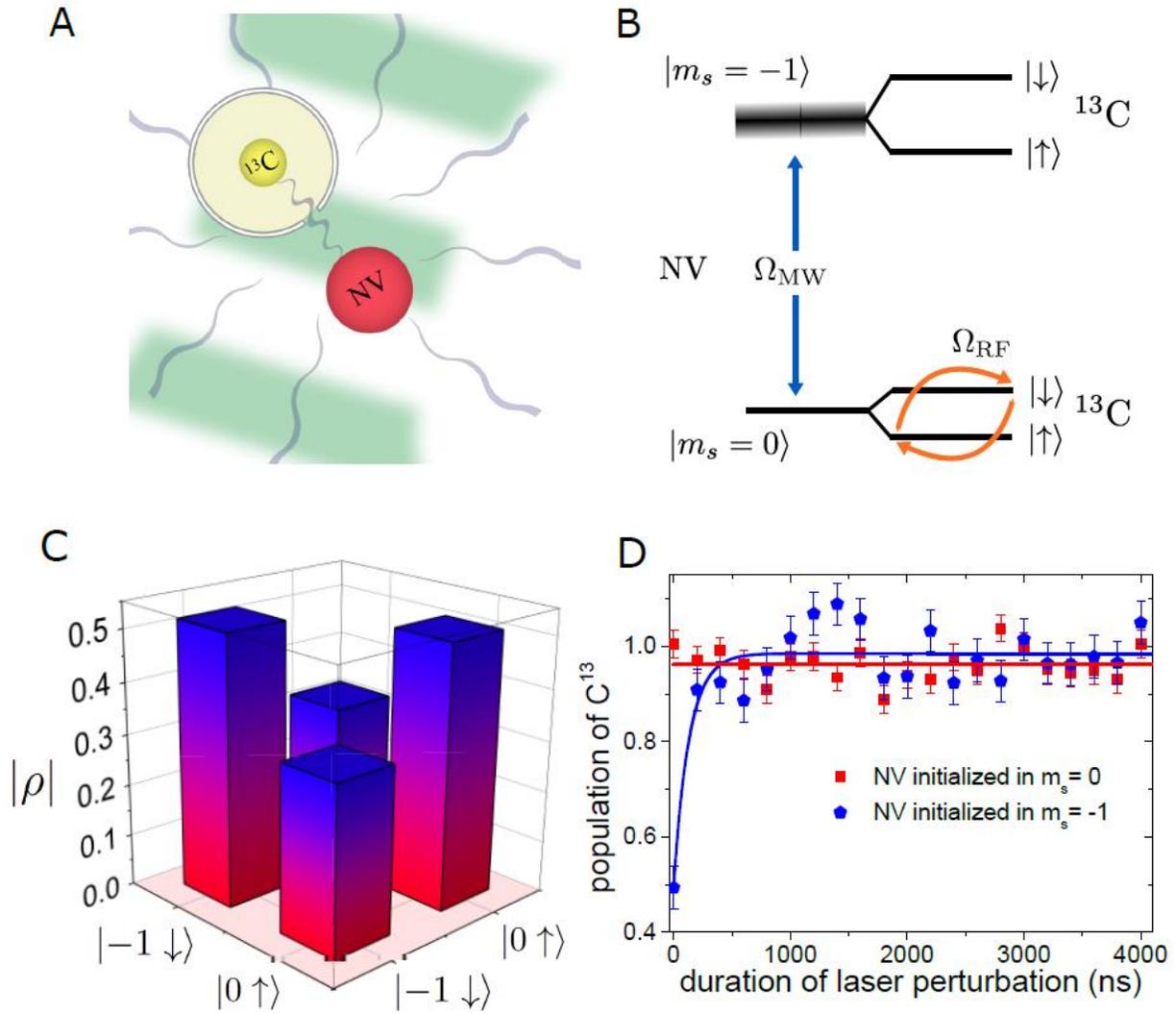

**Fig. 1**. Quantum system of interest and its fundamental properties. **A.** Quantum register formed by a NV and a near-axial $^{13}$C nucleus (auxiliary qubit) in the diamond lattice. **B.** Energy level structure of the quantum register, showing the ground state spin manifold $|0\rangle$, $|-1\rangle$ of the NV center and the spin manifold $|\uparrow\rangle$, $|\downarrow\rangle$ of the $^{13}$C spin. The NV-spin transitions are driven by a resonant microwave signal $\Omega_{MW}$ and the nuclear spin is driven by a radiofrequency signal $\Omega_{RF}$. To resolve the energy degeneracy of the NV spin states $|\pm 1\rangle$ and to define the NV qubit corresponding to the states $|0\rangle$ and $|-1\rangle$, an external magnetic field of about 340 G was applied. The magnetic field is aligned to the symmetry axis of the NV center. The coupling between NV center and nuclear spin is given by the hyperfine interaction, which in our case shows only a parallel component of 50 kHz (note that other non-axial nuclear spins can also be employed as robust qubits). The depicted broadening of the $|-1\rangle$ energy level corresponds to fluctuations of the NV Larmor frequency induced by the environment. **C**. Quantum state tomography of the prepared state $|+\rangle|\uparrow\rangle+|-\rangle|\downarrow\rangle$, showing a fidelity of 80% to create necessary entanglement between electronic and nuclear part of register. The absolute values of the entries of the density matrix $\rho$ are shown. **D**. Nuclear coherence under optical excitation of the NV center as measured by free induction decay of the nuclear spin. During the nuclear free evolution time, continuous optical excitation of NV is performed. The NV was initially prepared in either the $|0\rangle$ or the $|-1\rangle$ state. After free evolution of the nuclear spin, a $\pi/2$ pulse conditional on the electronic spin state $|0\rangle$ was applied to the nuclear spin to convert the nuclear coherence to population difference and a projective measurement was subsequently performed.

**Figure 2.**

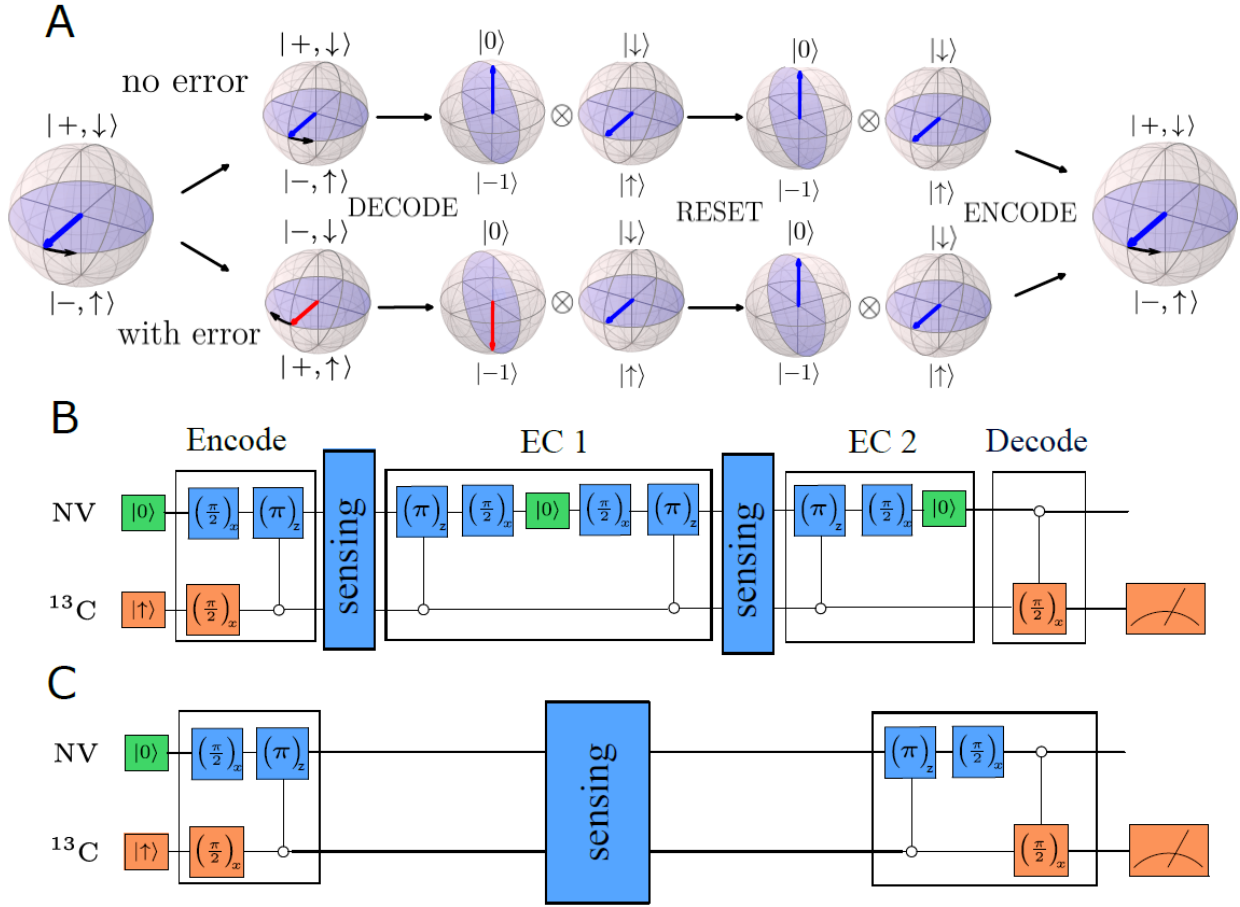

**Fig. 2.** Temporal evolution of the hybrid register and the implemented quantum sensing with error correction (QSEC) scheme. **A**. QSEC protocol in Bloch sphere representation, showing decoding (disentanglement), reset of the NV spin, and encoding. The error in this scheme is a single phase-flip of the NV. No time evolution between error and correction steps is assumed. The presented scheme is the basic correction step, which is applied repetitively (up to two times). The phase of the initial state corresponds to the signal. **B**. Measurement sequence used in the experiment when QSEC is used. First the electron and the nuclear spins were initialized via a laser pulse and single shot measurement. To encode the system in the $|+\uparrow\rangle+|-\downarrow\rangle$ state, local operations are applied with microwave and radiofrequency pulses. For entanglement creation with a controlled-not gate ($CNOT_z$), the parallel hyperfine interaction between NV and $^{13}C$ is used. All control pulses are performed on resonance to a single NV transition corresponding to a certain nuclear spin state. After encoding, a microwave signal with variable time duration is applied, resonant with the electronic part of the register and coherent signal accumulation occurs. During the first error correction step, entanglement between the NV and nuclear spins is removed by a CNOT gate, then a reset of the NV is performed and finally the entangled state is prepared for further sensing. After the second period of coherent interaction with the microwave signal, a further error correction step is applied. To readout the phase of the state $|+\uparrow\rangle + e^{i\varphi}|-\downarrow\rangle$, it is mapped onto the local phase of the nuclear spin and readout after a conditional $\pi/2$ pulse by performing a nondemolition measurement. **C.** The measurement sequence used in the experiment when no error correction is used. The measurement sequence follows that shown in B, but without the intermediate and last error correction steps.

**Figure 3.**

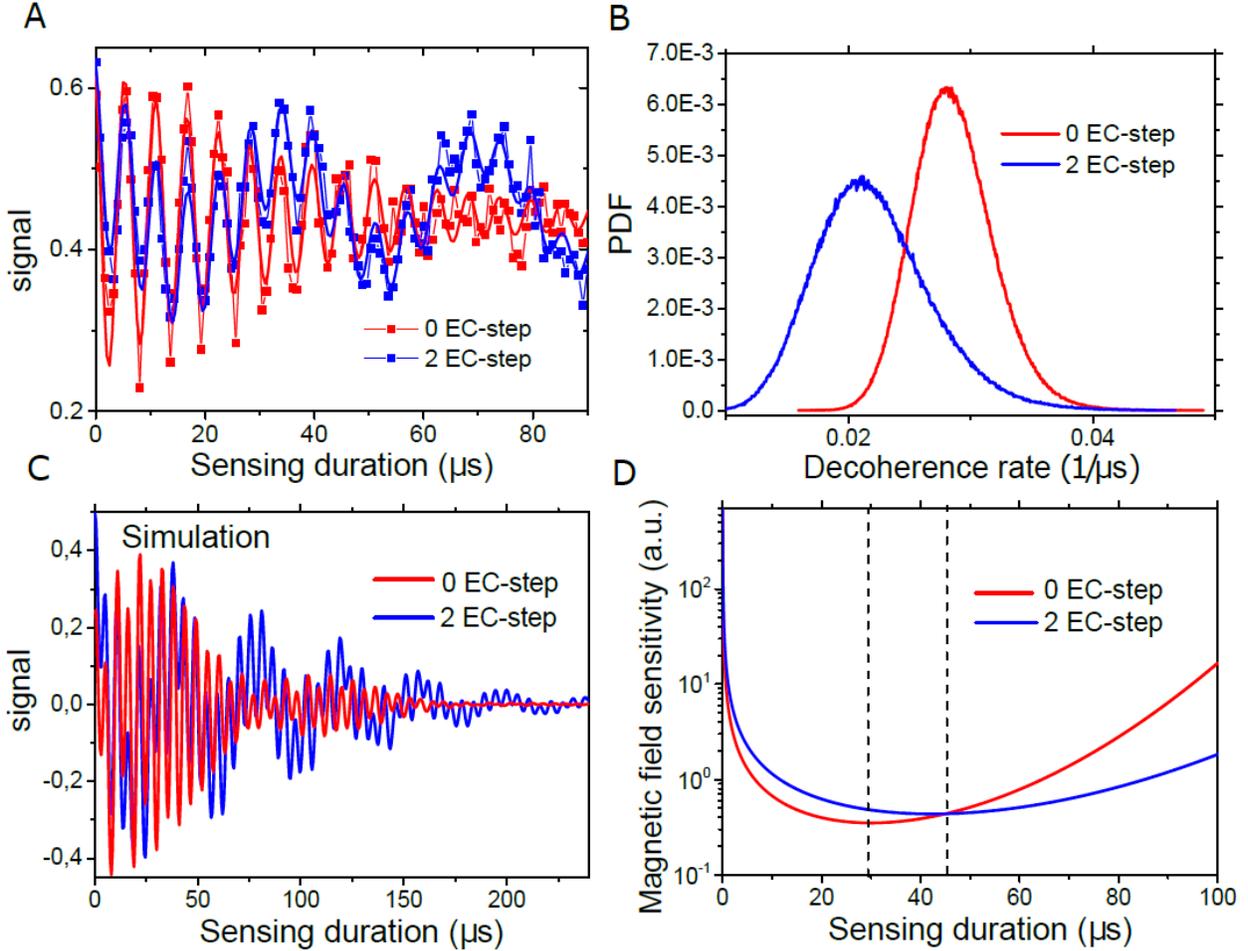

**Fig. 3.** Performance of QSEC against natural noise and sensitivity analysis. **A**. NV Rabi oscillations based on the sequences in Figure 2.B and in 2.C comparing the signal when no EC is used and when two EC steps are performed. In the case where EC was used, a slower oscillation appears. This oscillation corresponds to half of the hyperfine interaction strength ($50 kHz$) and is connected to the unitary evolution mediated by the hyperfine coupling within the register itself. For both experiments, the same overall measurement time was used. The solid lines correspond to fits. In the case of no EC, we used $f(x) = A \cdot e^{-\gamma \cdot x} \cdot \cos(\omega \cdot x + \varphi) + c$ and in the case with EC $f(x) = A_1 e^{-\Gamma \cdot x} \cos(\omega_1 \cdot x + \varphi_1) + A_2 e^{-\gamma \cdot x} \cdot \cos(\omega_2 \cdot x + \varphi_2) + c$ to fit the measurement data. **B**. Probability distribution function (PDF) of the fit parameter $\gamma$ based on Bayesian estimation. The fit parameter $\gamma$ corresponds to the rate of the observed decoherence with (blue) and without (red) QSEC. **C**. Simulation of the measurement data, based on a master equation which takes slow fluctuation of the NV Larmor frequency into account. More information about the simulation model can be found in [18]. **D**. Magnetic sensitivity of the sensor as a function of sensing time with and without QSEC. The shown sensitivities are calculated based on the estimated parameter describing the measurement data shown in Figure 3.A. The dashed lines indicate the sensing duration which gives the best achievable sensitivity. The sensing duration is the time of sensing, neglecting the time needed for initialization, readout and error correction.

# Supplemental Material

## Experimental information

For optical manipulation and detection of single nitrogen vacancy (NV) center in bulk diamond we used a home-built confocal microscope. Initialization and readout of the NV ground state spin was performed with green laser excitation (532nm) and detection of the corresponding fluorescence with an avalanche photodiode. Static magnetic fields were applied with permanent magnets. The electronic (NV) as well as nuclear ($^{13}$C) spin transitions were driven by microwave and radiofrequency fields applied using a thin copper wire located close to the NV center and generated by an arbitrary-waveform generator. The rotation axis was defined by the microwave phase. Rotations around the y-axis were performed by inserting a $\pi/2$ phase-shift to the microwave. The diamond sample was grown using chemical vapor deposition on a IIa diamond substrate. The isotopic composition of the overgrown layer was 99.9% $^{12}$C. Native (un-implanted) NV centers were used as qubits. This concentration of carbon isotope carrying nuclear spins (0.1 % $^{13}$C) allow to reach long phase memory time and to have a significant probability to find weakly coupled NVs in the frozen core.

## Readout and Initialization of the nuclear spin

To initialize and readout the nuclear spin, we use a single-shot readout [29] technique. In Fig. S.1.C the corresponding measurement sequence is shown. First, the NV center spin is polarized by a $300 ns$ laser pulse. Second, we transfer the state of the nuclear spin to the NV center spin with the help of a CNOT-gate (see Fig. S.1.D). In the end we perform a population measurement on the NV spin. This sequence is repeated 10000-times and the corresponding results of the NV spin readout are accumulated. By repeated application of the single-shot sequence, quantum jumps of the $^{13}$C are visible in the nuclear spin-dependent fluorescence time trace of the NV center (see Fig. S.1.A). In the corresponding histogram, shown in Fig. S.1.B, the two states of the $^{13}$C are visible. By setting a fluorescence threshold between the maximum of the two states, the nucleus can be initialized into a given state based on the measurement outcome. By measuring the probability for detecting the state which was originally prepared and assuming a perfect preparation, we observe a readout fidelity of 98% for the nuclear spin and a fidelity of initialization above 99% when a low initialization threshold is used.

## Simulation of Rabi oscillations with phase correction

The simulations in Figure 3.C are based on the following master equation $\rho' = -i[H_0 + H_C, \rho] + \gamma(t)\left(S_z \rho S_z - \frac{1}{2}\{S_z^+ S_z, \rho\}\right).$ With $H_0 = \Delta_{NV} S_z + A S_z I_z + \Delta_{13C} I_z$ and $H_C = x(t) S_x + y(t) S_y + X(t) I_x$. $\Delta_{NV} = -A/2 = 25 kHz$ is the detuning of the microwave, which is used for NV manipulation. $\Delta_{13C} = -A/2 = 25 kHz$ is the detuning of the radiofrequency, which is used for $^{13}C$ manipulation. $A$ is the hyperfine coupling and $S_z, S_x, S_y, I_z, I_x$ are the corresponding spin operators of NV and $^{13}C$. The functions $x(t), y(t), X(t)$ define the applied pulses (around x- and y-axis) corresponding to the measurement sequence described in the main text. $\gamma(t) = \frac{t}{T^2}$ is the decay-function and simulates slow Gaussian dephasing of the NV. We have chosen $T = 40 \mu s$, which is in good agreement with the experimental conditions.

## Simulation of QSEC when no error is present

In Fig. S.2 we present simulations when no error was introduced. The simulated protocol is based on the sequence shown in Figure 2 of the main text. The slower oscillation, visible when error correction is used, corresponds to the hyperfine interaction within the register. It is $25\ kHz$, which is half of the parallel component of the hyperfine interaction between NV and $^{13}C$.

## Simulation of QSEC in presence of discrete error in time

By introducing a full phase flip on NV in front of the intermediate error correction step, the simulations show full recovery of the signal when error correction is used (see Fig. S.3). In the case of no error correction, the signal is completely destroyed.

## Experimental correction of single phase error

In Figure S.4 we show experimental data when a discrete error in time was inserted. The phase flip error ($\pi$-rotation around z-axis) was introduced before the intermediate error correction step and was created by an off resonant $2\pi$-rotation around the x-axis the of the NV center. We observe a clear recovery of the signal when error correction was used.

# Experimental correction of bit-flip error

In this experiment we correct an artificially applied bit-flip error on the NV center. Artificial noise is used, since for the deep, well-protected NV center used in this experiment, bit-flip errors play a minor role compared to phase-flip errors. It is an example of fast noise and shows the complementary relationship between standard DD and QSEC protocols. Our measurement sequence is presented in Figure S.7.A. A more detailed protocol, showing implementation of EC, encoding and decoding is explained in Figure S.5. The sensing sequence is based on a Carr-Purcell-Meiboom-Gill (CPMG) sequence to sense a weak non-resonant applied AC-signal (phase-locked) with a frequency of $100 kHz$. The pulse spacing in time of the CPMG sequence was fixed and tuned to the signal resonance. In the last free evolution time of the CPMG, a single error ($R_y(\theta)$; rotation around $y$-axis with rotation angle $\theta$) was applied by strong resonant microwave pulses. The strength of the error ($\theta$) was varied by changing the duration of the resonant microwave pulse. A strong error corresponds to a NV $\pi$-pulse and is performed in approximately $60 ns$. After the CPMG sequence, the EC step was applied. We performed the experiments with different error strength and delay time between error and error correction. The measured sensitivities are shown in Figure S.6. In the case of two EC, the intermediate sequence (CPMG+EC, Figure S.7.A) was repeated two times. In the experiment shown in Figure S.6.A we varied the error strength and used no delay time between error and correction. In the experiment presented in Figure S.6.B we used a full error ($\pi$-pulse) and varied the delay time between error and correction. In the experiment presented in Fig. S.7 we averaged the observed signals corresponding to multiple experiments where a single error with different delay time and strength (rotation angle) was applied. For the strength of the error we have chosen one out of the rotation angles ($0.5\pi, 0.75\pi, 1.0\pi$) and for the delay time one out of ($0\mu s, 0.6\mu s, 1.2\mu s, 1.8\mu s$). The corresponding sensitivity is presented in S.7.C.

**Figure S1.**

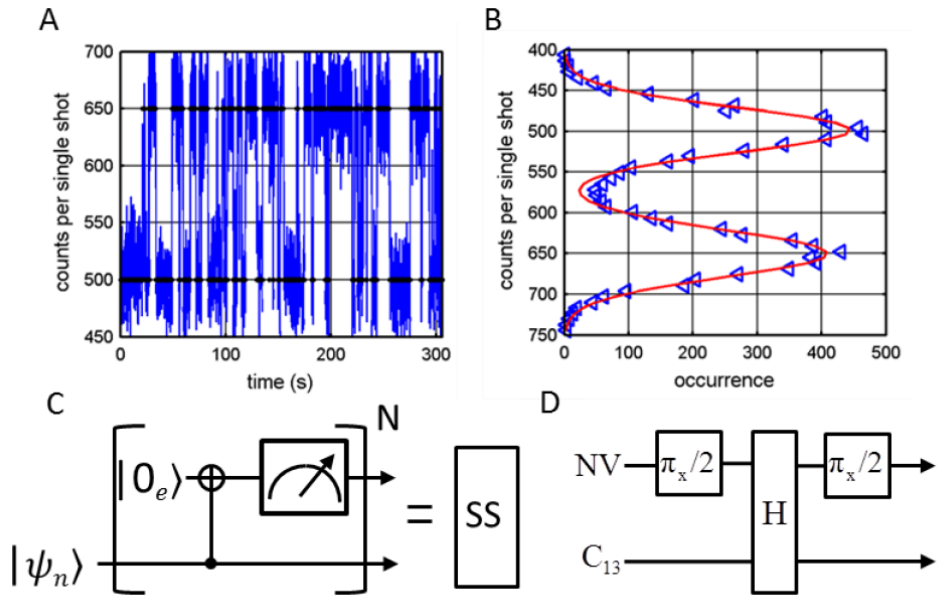

**Fig. S1. A:** Single-Shot readout of the $^{13}$C spin. Shown is the nuclear spin dependent NV fluorescence over time. Each point corresponds to an accumulated fluorescence when the sequence presented in C is 10.000 times repeated. **B**: Corresponding histogram of the spin-dependent fluorescence. **C:** Sequence of a single-shot measurement of the nuclear spin. **D:** Sequence of our used CNOT-gate. Between two resonant $\pi/2$-gates applied to the NV center, there is a free evolution time of half the period of the hyperfine coupling period.

**Figure S2.**

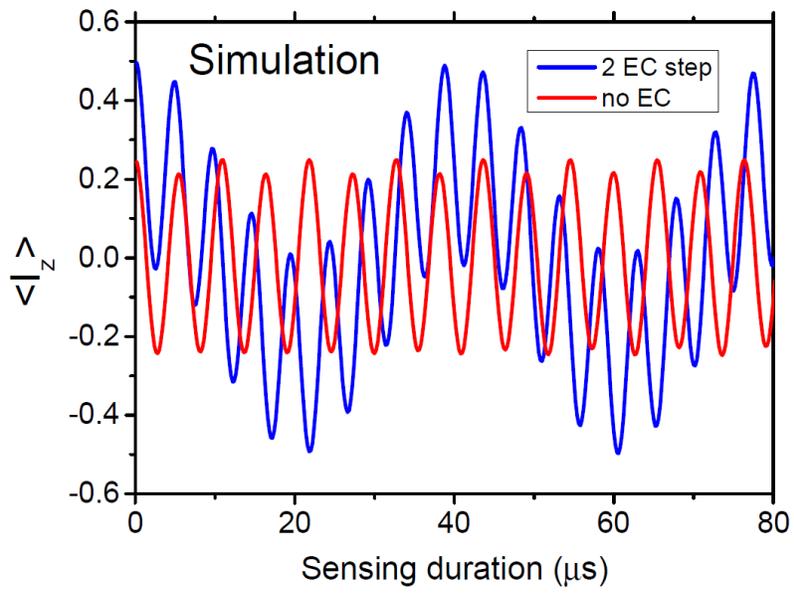

**Fig. S2.** Simulation of the measurement protocol presented in the main text. In this case no noise is introduced, only a resonant driving of the NV.

**Figure S3.**

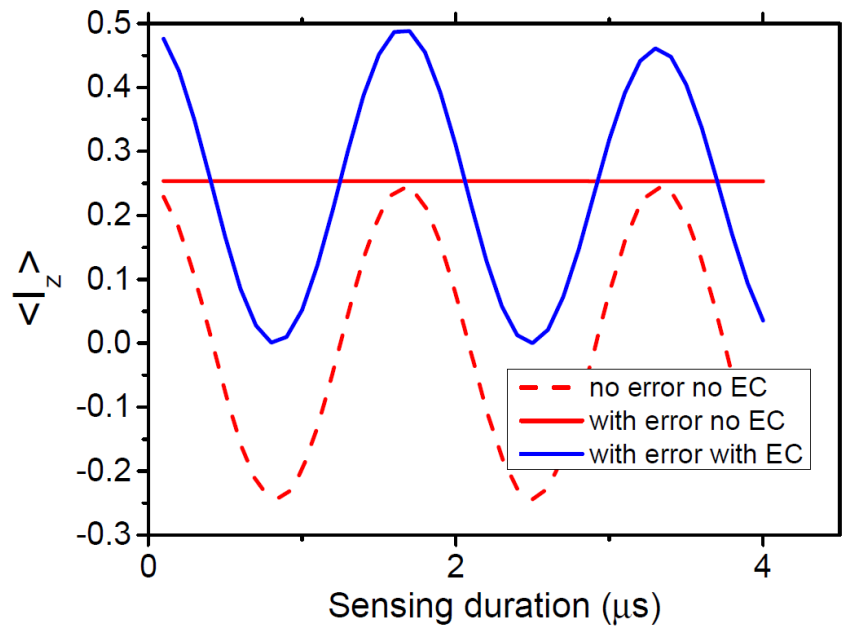

**Fig. S3.** Simulation of the measurement protocol presented in the main text when in front of the intermediate EC step a full phase flip error is introduced or not. In the case of no EC, the error was introduced in the middle of the sensing period.

**Figure S4.**

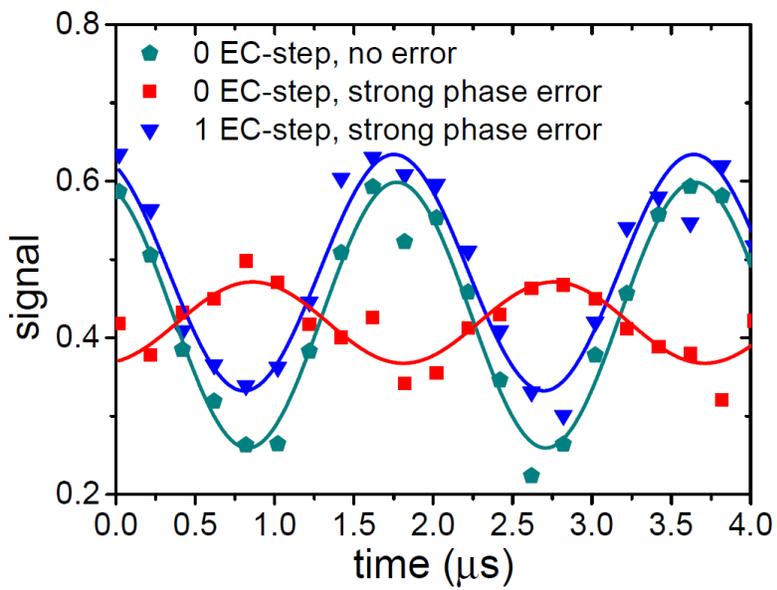

**Fig. S4.** Experimental correction of single phase-flip error.

**Figure S5.**

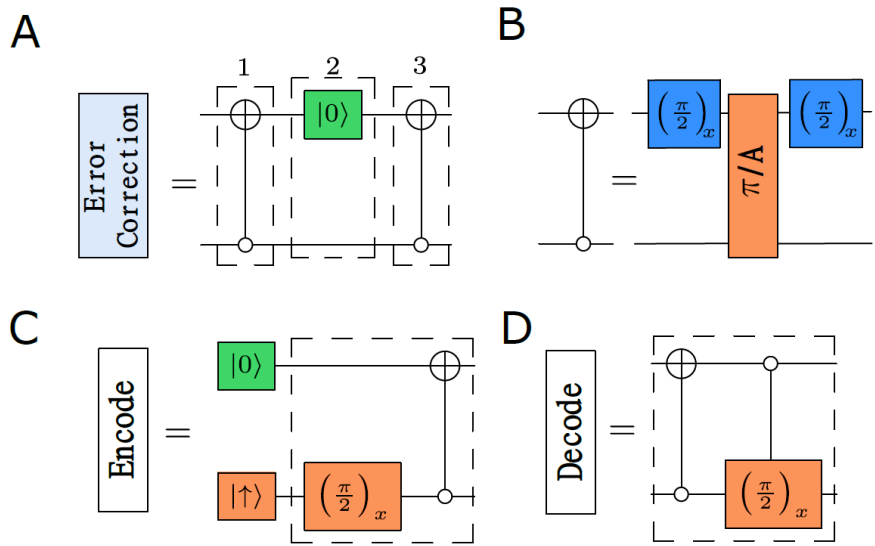

**Fig. S5.** Measurement protocol in the case of bit-flip error. **A.** Sequence of the error correction step. In between two CNOT gates a reset of the NV is performed. **B.** Sequence of the CNOT gate. **C.** Protocol to encode the register. **D.** Protocol to decode the register.

**Figure S6.**

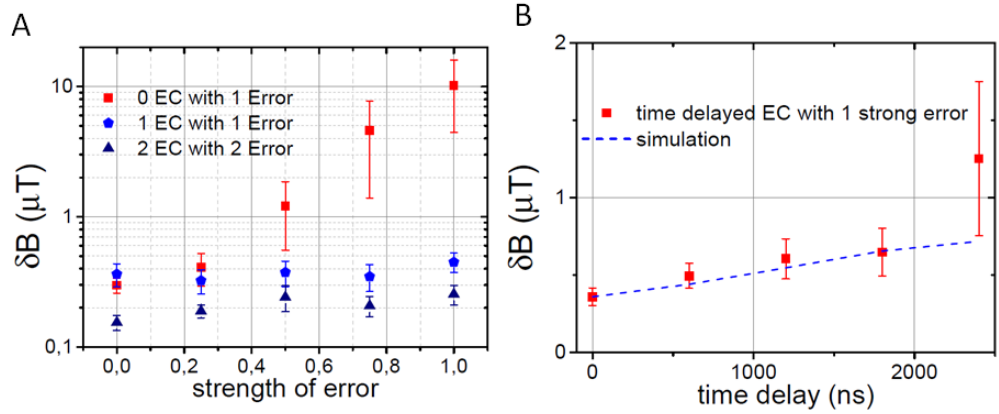

**Fig. S6.** Sensitivity of the magnetometer, when a single bit-flip error with different strength (A) and delay time (B) between error and correction step is applied.

**Figure S.7**

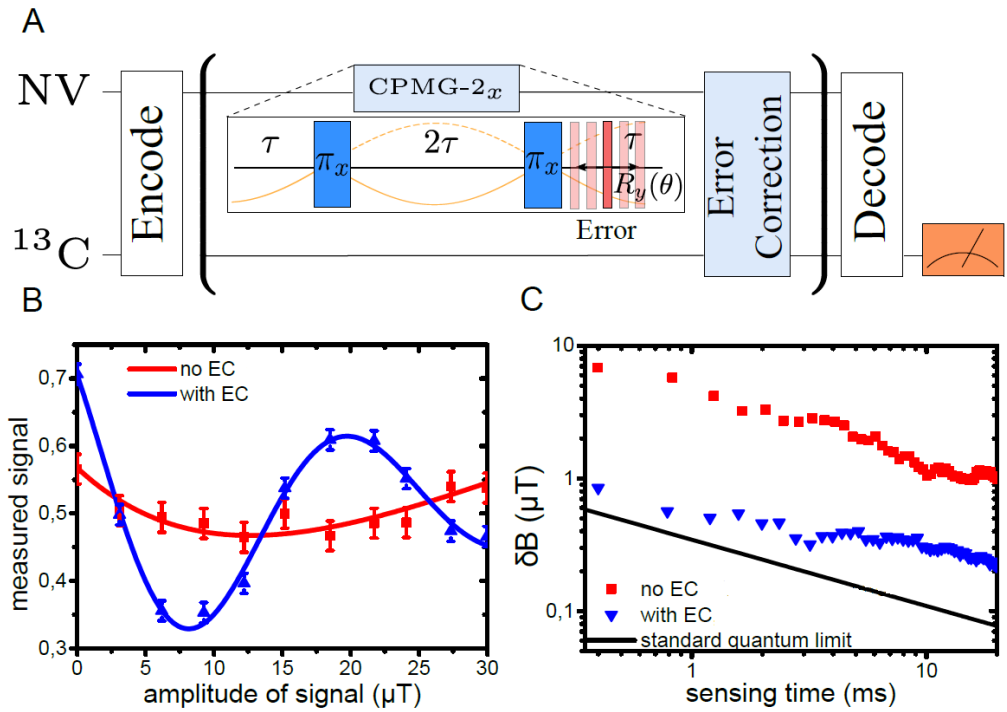

**Fig. S7. A.** Measurement sequence in the case of variable bit-flip error in time and in strength. **B.** Averaged Signal when different error strength ($0.5\pi, 0.75\pi, 1.0\pi$) and different delay time ($0\mu s, 0.6\mu s, 1.2\mu s, 1.8\mu s$) between error and correction step are used. **C.** Corresponding sensitivity. The sensing time is the overall measurement time without the duration of the initialization-, readout- and EC-step.